# Metallized Film Capacitor Lifetime Evaluation and Failure Mode Analysis

*R. Gallay*
Garmanage, Farvagny-le-Petit, Switzerland

**Abstract**
One of the main concerns for power electronic engineers regarding capacitors is to predict their remaining lifetime in order to anticipate costly failures or system unavailability. This may be achieved using a Weibull statistical law combined with acceleration factors for the temperature, the voltage, and the humidity. This paper discusses the different capacitor failure modes and their effects and consequences.

## 1   Capacitor technologies

The following different power capacitor technologies are used in inverters:

– Electrolytic capacitors characterized by very big capacitance per volume unit, but with low rated voltages and very important power losses due to the ionic conductivity. In particular, the bigger the capacitance density, the lower the rated voltage.

– Film foil capacitors made of dielectric films between two plain aluminium foils. These capacitors can sustain very high currents.

– Metallized film capacitors, which are made with dielectric films with a metallic coating on the surface. With this technology the electric-field stress may be much bigger than with film capacitors thanks to the metallization self-healing capability.

Today the dielectric films that are used are mainly polypropylene (PP) or polyethylene terephthalate (PET). Formerly, paper (PA) was used in film foil technology—either pure paper or mixed with polypropylene (DM). In special applications, where high temperatures are required, polyethylene naphthalene (PEN) up to 125°C or polyphenylene sulfide (PPS) up to 150°C are used.

PET presents the following advantages over PP: a dielectric constant 50% bigger ($\varepsilon = 3.3$ versus 2.2), which means 50% more capacitance in the same volume, a better mechanical resistance (which means a higher endurance to self-healing), and the possibility of manipulating thinner films, consequently leading to a smaller capacitance and a higher exploitation temperature (+10°C). The negative point is that the loss factor is ten times larger, which means a ten-fold increase in temperature elevation for the same rated power. The nominal electrical field is about the same.

The capacitive elements must be dried to remove moisture, which would cause accelerated aging and bigger losses if left in the capacitor. In the case of power capacitors, the dried elements are either impregnated with vegetable oil or with gas ($SF_6$, $N_2$, etc.).

The dielectric films are either wound or stacked before being inserted in a plastic or metallic container. The best winding machines are required to produce active wound elements of reliable quality in the case of oil-free capacitors. One way of overcoming the difficulty of controlling the space ratio between the gas and the film in the winding curves is to wind the film on a large-diameter wheel and to cut the film layers to obtain a stack.

The plastic containers are not completely moisture tight—there is always some residual permeability in polymers. In the case of metallized films, this may lead to electrode corrosion when the capacitors are submitted to environmental conditions of high humidity.

The electric-field stess in metallized film capacitors may be much larger than in film foil capacitors. This is obtained thanks to the ability of the electrodes to self-heal. If a breakdown occurs in the polymer, the current will increase through the defect and on the electrode near the defect. Close to the defect the current density will be big enough to evaporate the 100 nm metallic layer. If the capacitor is well designed, the phenomenon will stop when the diameter is large enough to insulate the defect and small enough not to damage the film. The electrode resistance (given in ohm/square) is the key parameter to define to achieve good self-healing behaviour, with Joule losses as small as possible. A thick metallized layer will present a lower resistance, but higher energies will be involved during the self-healing process, leading to greater damage [1-5].

## 2 Capacitor failure modes

Most of the metallized film capacitors fail because the capacitance drops below the required tolerance. This normally occurs after the expected lifetime given by the manufacturer. The capacitance drop is generally accompanied by an increase of the loss factor.

From a general point of view, the causes of capacitor failures may occur because of bad design, bad processes, or inappropriate application conditions. During the design phase, the following causes may lead to failure: the dielectric film is too thin, insulation distances are too small, the metallization layer is too thick or too thin, or the conductor is the wrong size. During production, causes may include the following: poor mechanical tension control during the winding, bad drying (leaving too high a humidity content in the capacitor), or bad sealing. In application, the causes may be: higher voltages, EMI, lightning, higher temperature, or a high humidity environment.

The failure modes are a little more complicated to describe because different causes may lead to the same modes. Figure 1 gives a non-exhaustive summary of the possible failure modes which can occur in metallized film capacitors.

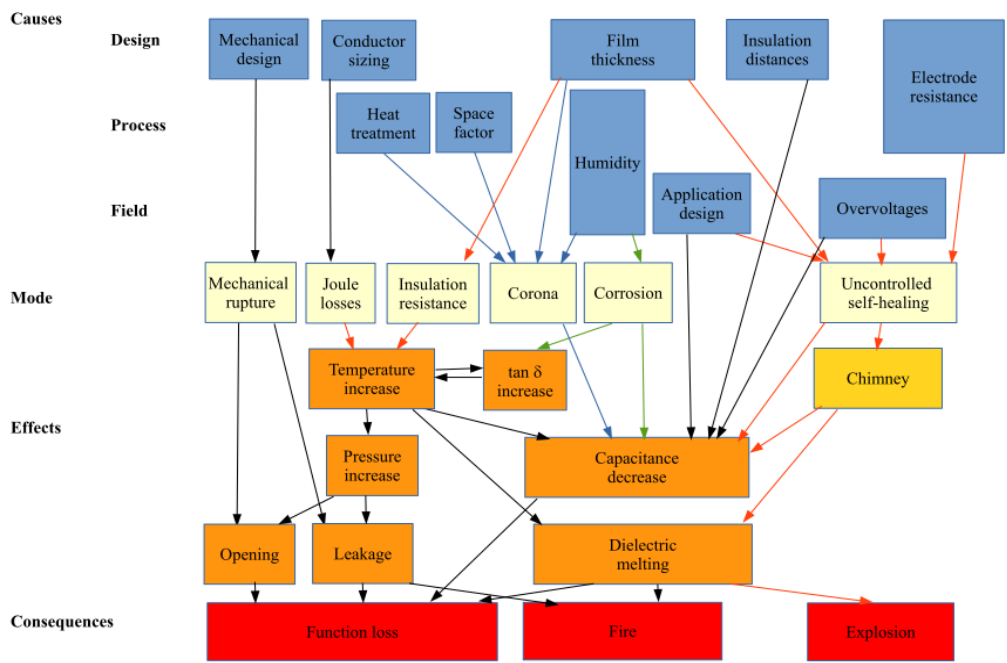

**Fig. 1:** Metallized film capacitor failure modes with their causes, effects, and consequences

For example, bad space factor control of the dielectric films during the winding operation will be the cause of the electrode corona demetallization, which will lead to a fast capacitance drop and to the loss of functionality of the capacitor.

A bad choice of the metallization resistance value, or poor metallization control during the film manufacturing process, leads to bad self-healing management, which may damage the dielectric film mechanically and produces heat which is transmitted locally to the next film layers. At this location the dielectric strength of the film drops and breakdown may occur. Consequently, chimneys of melted polypropylene may appear through the winding. The formed channel is conductive, inducing a drop in the insulation resistance and a leakage current that can generate enough heat to melt the polypropylene and increase the internal pressure of the capacitor. Along with bad metallization resistance, the final consequence can, in the worst case, lead to fire ignition or even a capacitor explosion.

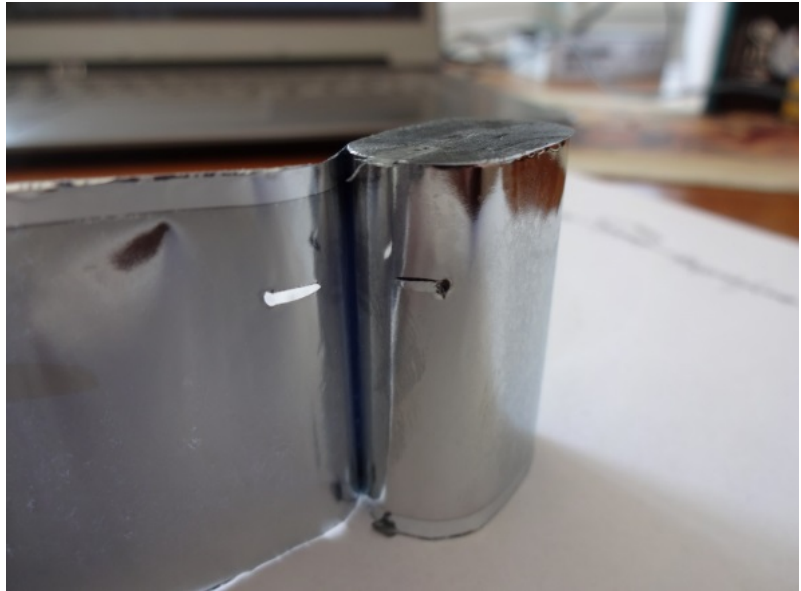

**Fig. 2**: Chimney through the film layers in the capacitor winding

One of the main failure modes is often due to high currents, which increase the capacitor temperature, leading to a reduction of the breakdown voltage and, in the worse cases, even melting of the capacitor. In this regard, the shape of the capacitor is very important. For high-power applications, it is important to build short elements in order to reduce the current path length and increase the number of parallel layers, and consequently reduce the heating. The current capability of a capacitor is specified through the series resistance $R_s$ and the loss factor $\tan \delta_s$ at different frequencies. The relation between the two factors, in the high-frequency domain where the effect of the insulation resistance is negligible, is given by the linear relation

$$\tan \delta_s = \frac{Z_R}{Z_i} = R_s \omega C, \tag{1}$$

where $C$ is the capacitance, $\omega = 2\pi f$ is the frequency, $Z_R$ is the impedance real part and $Z_i$ the impedance imaginary part.

The presence of humidity in the capacitor, because of poor drying during the manufacturing process, or because the moisture permeability of the material was too high, or because the humidity level where the capacitors are installed was too high, may lead to three failure modes with different effects and consequences.

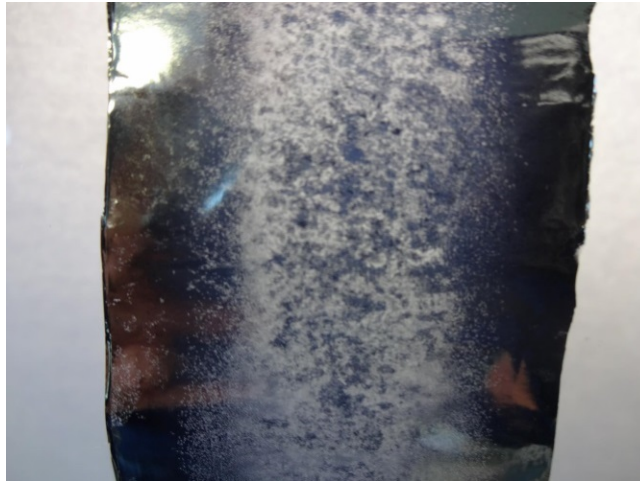
**Fig. 3**: Electrode corrosion due to the presence of moisture

The first is electrode corrosion (see Fig. 3) [6-8], where the series resistance will slowly increase over time. The effect is a loss factor increase due to the electrode thickness reduction and a heat dissipation increase. The elevation of the temperature will accelerate the capacitance loss because of the reduction of the dielectric strength with temperature, ending with the loss of functionality of the capacitor.

The second effect (see Fig. 4), today known as 'corona' [9-11], is due either to a decrease of the dielectric strength of the gas present in the capacitor in the gaps between the dielectric films or to a poor space factor control of the films. The bigger the gap, the more severe the problem. The thickness of the gap is characterized by the space factor, which is the ratio of the dielectric thickness to the total distance between the electrodes. This space factor is very difficult to control in curves of flat windings, leading manufacturers to build either round winding or stacks. Only performant winding machines can achieve good space factor control by managing the mechanical tension of the films during the winding. The consequence of this is a fast capacitance decrease due to the appearance of corona discharges on the electrode edges, i.e., the locations where the electrical field is more intense due to the point effect. In the case of segmented metallization, the corona failure mode may also propagate from the non-metallic lines which separate the active electrode metallic areas.

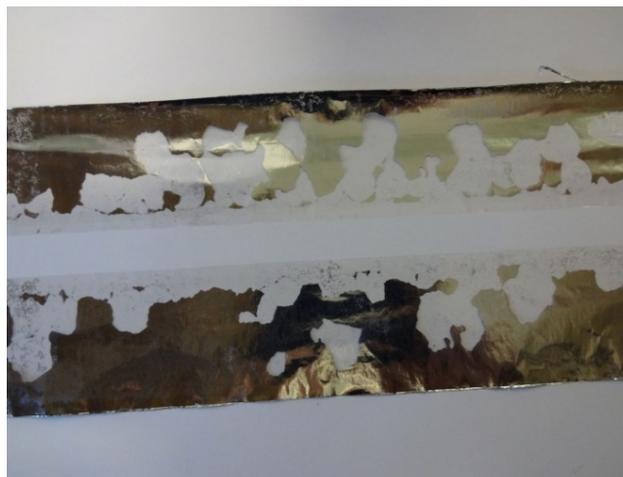
**Fig. 4:** Demetallized electrodes by corona arcing in the gas gap between the films

The third failure mode is a reduction of the insulation resistance, which is the parallel resistance of the capacitor. A decrease in insulation resistance leads to an increase in current leakage from one electrode to the other. This phenomenon is present at low frequency. It may be measured via either the loss factor (tan $\delta$) or the d.c. resistance $R_p$. The relation between the two parameters is given by the following relation (only true at very low frequencies):

$$\tan \delta_p = \frac{Z_R}{Z_i} = \frac{1}{R_p \omega C}. \qquad (2)$$

This later failure mode may have a runaway behaviour. The more the insulation resistance decreases, the more heat is produced, and the more the temperature increases, which leads to a new insulation decrease. This phenomenon may end with the appearance of chimneys and melting of the dielectric.

## 3   Lifetime expectancy

The lifetime [12] of a capacitor is the time to failure, where failure is defined as the lack of ability of a component to fulfil its specified function. The failure modes are classified into two main categories: 'early failures' and 'wear out failures', which are reflected in the curve known as the 'bathtub' curve (Fig. 5): at the beginning of the component's existence, in its 'infancy', the failure rate is rapidly decreasing. These 'youth' failures are normally screened by routine tests performed by the manufacturer. They are due to design and process weaknesses which have not been detected by the design and process failure modes and effects analysis FMEA performed during the development. They are more probably due to production process variations or to changes in material quality. The process variations are due to tool wear, operator change, and lack of formation. This early failure mode is not taken into account by the Weibull model theory. In normal operation this failure process should not be observed in the field of applications. If it occurs, the capacitors are normally covered by manufacturer's product warranty.

Once the 'early failures' regime is past, the failure rate starts to follow a statistical prediction law which depends on several parameters that may be defined experimentally as a function of the voltage, the temperature, and the environmental humidity. It has been shown that a Weibull statistic can provide a good prediction of the capacitor lifetime expectancy.

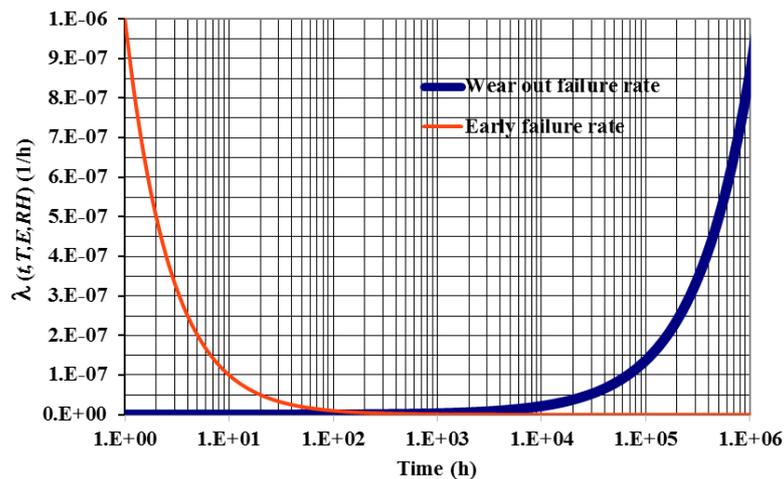

**Fig. 5:** Bathtub curve of the failure rate function showing the infancy or early failures occurring at the beginning of the component's life (in red) and the wear out curve (in blue) which is defined by aWeibull law with 2 parameters: the power factor $p = 1.8$ and the inverse of the time necessary for 63% of the sample to fail $\lambda_0 = 1/1{,}500{,}000$ h$^{-1}$.

The failure rate $\lambda(t)$ may be given in FIT (failure in time), which is the number of failures occurring during $10^9$ h of working of one object, i.e., $3\text{E}^{-7}$ h$^{-1}$ corresponds to 300 FIT. The Weibull failure rate is given by

$$\lambda(t) = \lambda_0 \, p \, (\lambda_0 t)^{p-1}, \tag{3}$$

where $\lambda_0$ must not be confused with $\lambda(t)$. $\lambda_0$ is a constant (independent of time, but dependent on temperature, voltage, and humidity) which corresponds to the inverse of the time necessary for 63% of the sample to fail, and $\lambda(t)$ is the inverse of the mean time to failure (MTTF). In the wear out failure region, $\lambda(t)$ increases with time. The manufacturer specifications give the maximum value of $\lambda(t)$ within the announced lifetime: for example, 150 FIT and 100,000 h of lifetime expectancy in Fig. 5. The slope parameter of the Weibull law is denoted by $p$.

The survivor or Weibull reliability function $R(t)$ is the probability that a capacitor has not failed or has not lost its function at time $t$ and is still working. The survivor function is given by

$$R(t) = e^{-(\lambda_0 t)^p}. \tag{4}$$

When multiplied by the number of capacitors $N$ in the batch, this gives the expected number of capacitors still working after time $t$. There are capacitor manufacturers [13] that use a simple exponential model instead of the more complex Weibull model. Actually the exponential model corresponds to a Weibull model where $p = 1$. In this exponential model the failure rate is constant over the time $\lambda(t) = \lambda_0$. To fit this model to the actual statistical behaviour of capacitors, manufacturers limit the exponential model to a time period which they call the 'service life of the product'. After that time period the failure rate starts to increase.

Weibull statistics can also be used to predict the capacitance evolution of a metallized capacitor under electrical, thermal, and humidity stresses. In such cases, the failure definition will be, for example, 1% or 1‰ capacitance loss, depending on the available resolution of the measurement device. The capacitance will be given straightforwardly by the survivor function. The Weibull reliable life, which gives the expected lifetime of a capacitor for a given reliability level (the proportion of remaining working objects, in our case capacitance) is

$$T_R = \frac{1}{\lambda_0}\{-\ln(R)\}^{1/p}, \tag{5}$$

where $\lambda_0$ is the failure rate for the special case when $1/e$, or 36.8%, of the samples still remain.

**Table 1:** Capacitor lifetime expectancy factor as a function of the required capacitance minimum in an exponential model.

| Reliability (%) | Lifetime, $1/\lambda_0$ (h) |
|---|---|
| 36.8 | 1 |
| 50 | 0.693 |
| 63.2 | 0.500 |
| 80 | 0.223 |
| 90 | 0.105 |
| 95 | 0.051 |
| 98 | 0.020 |

If the manufacturer gives a capacitor failure rate of 50 FIT at 40°C and $U_n/2$ for an exponential model, it means that the lifetime expectancy for a capacitance drop tolerance of 10% will be $2.1 \times 10^6$ h in these conditions.

## 4 Aging acceleration factors

The speed of capacitance drop depends on the temperature, the voltage, and the humidity. The increases in these parameters are considered as aging acceleration factors. These factors are determined experimentally based on the following theories.

### 4.1 Temperature

It has been shown [14] that capacitor aging as a function of the temperature follows an Arrhenius law, in other words an exponential law

$$t(T) = t_{T_n} \exp\left(\frac{E_a}{k_B}\left(\frac{1}{T} - \frac{1}{T_n}\right)\right), \qquad (6)$$

where $t_{T_n}$ is the expected lifetime at a reference temperature, 70°C or 85°C for example, $k_B$ is the Boltzmann constant, and $E_a$ is an activation energy. A relatively good fit of Epcos/Vishay factors (see Fig. 6) may be obtained with a ratio $E_a/k_B$ = 7000 K [15]. Between 40°C and 70°C there is an acceleration factor of 7.1 with the parameters considered.

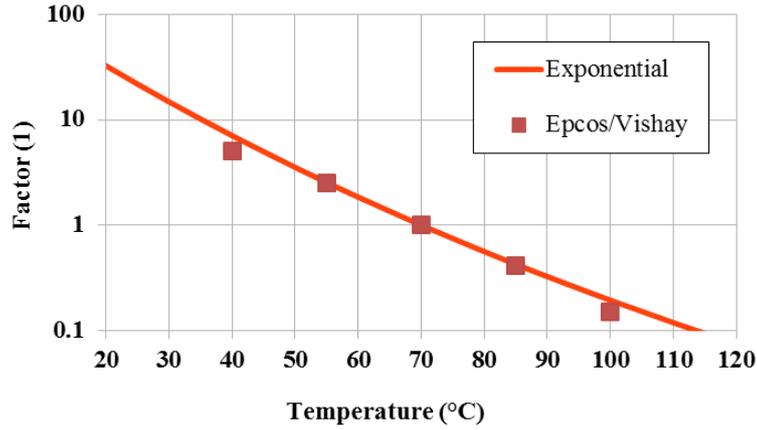

**Fig. 6:** Temperature acceleration factor

### 4.2 Voltage

Regarding voltage dependency, authors use either an inverse power law

$$t = t_{U_n}\left(\frac{U}{U_n}\right)^{-n} \qquad (7)$$

or an exponential law

$$t = t_{U_n} \exp\left(-\alpha \frac{(U - U_n)}{U_n}\right), \qquad (8)$$

where $t_{U_n}$ is the expected lifetime at the nominal voltage or reference voltage and $t/t_{U_n}$ is the voltage acceleration factor.

A careful examination shows that these laws do not differ much when considered between 0.7 and $1.3U_n$. To sketch Fig. 7, $n$ and $\alpha$ have been both set to 3.5.

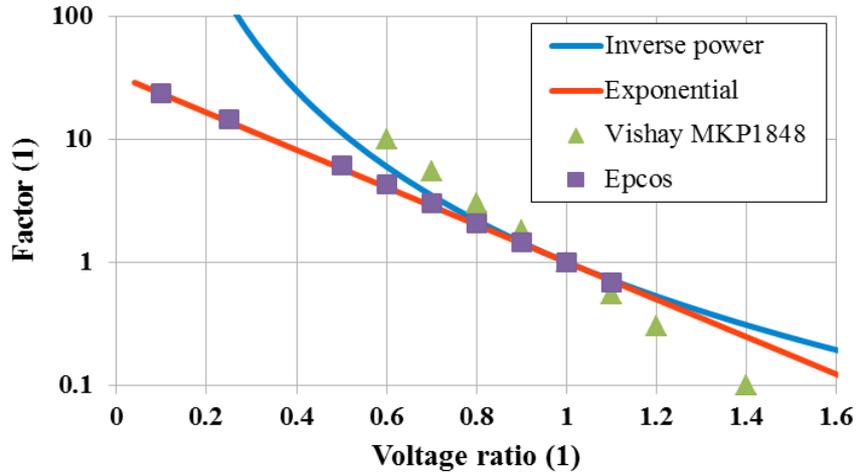

**Fig. 7:** Acceleration factor of the aging as a function of the voltage level. In this diagram the inverse power law and the exponential law are both parameterized with a factor 3.5. The Vishay data would fit better with an exponential law with factor 5 [16, 17].

Between $U_n/2$ and $U_n$ there is an acceleration factor of approximatively 5 (Epcos) to 10 (inverse power) with the parameters considered. The discrepancy between the values may be attributed to the different technologies produced by the different manufacturers. It is interesting to note that the Vishay display on p. 6 of Ref. [18] has almost the same accelerating factors as Epcos.

### 4.3 Humidity

Humidity is a concern for capacitors contained in plastic because moisture can permeate through this type of material. Once inside the capacitor, the moisture has several effects: first, it decreases the electrical strength of the gas in the case of oil-free capacitors, leading to corona demetallization of the electrode, and secondly it corrodes the electrode. When moisture is present in the dielectric film, the loss factor is increased, because of the presence of water dipoles, and the insulation resistance is reduced, leading to current leakage and generation of heat. The lifetime as a function of the humidity level can be estimated using the following relation [19]:

$$t(RH) = t_{H_n} \left( \frac{RH_n}{RH} \right)^m, \tag{9}$$

where $t_{H_n}$ is the expected lifetime at a reference humidity level.

### 4.4 Capacitor lifetime expectancy calculation

To design and size a capacitor correctly engineers must work through two steps. First, they must determine the law parameters: $E_a$ for the temperature, $n$ for the voltage, and $m$ for the humidity. This is basically achieved by defining a plan of the experiment with three different temperatures, voltages, and humidity levels. The second step is the calculation of the lifetime expectancy as a function of the customer specifications. For example, for a solar inverter, half of the time there is no voltage and the temperature is 20°C, 20% of the time the voltage is maximum and the temperature reaches 90°C, and 30% of the time the voltage is 80% of the maximum and the temperature is 60°C. Each stress condition must be converted to a reference condition value. The sum of the contributions will determine the lifetime expectation.

**Table 2:** Lifetime expectancy with indicative data

| Duration (%) | Voltage ($U_n$) | Temperature (°C) | Converted aging weight @ reference conditions(%) |
|---|---|---|---|
| 50 | 0 | 20 | 5 |
| 30 | 0.8 | 60 | 25 |
| 20 | 1 | 90 | 70 |

In most applications, in the first step the required dielectric thickness is calculated at the reference temperature for the specified voltage distribution. In the second step the temperature distribution is used to adapt the dielectric thickness to the temperature profile requirement. In the section 'Power cap sizing' of the site http://www.garmanage.com [20], there is a tool which allows the user to calculate the lifetime expectancy with ten different stresses.

Taking again the example of a failure rate of 50 FIT in an exponential model, which gives a lifetime expectancy of $2.1 \times 10^6$ h at 40°C and $U_n/2$, it may be calculated by multiplication of Eqs. (7), (8), and (9)

$$t(T,U,RH) = t_{T_n,U_n,RH_n} \exp\left(\frac{E_a}{k_B}\left(\frac{1}{T}-\frac{1}{T_n}\right)\right)\left(\frac{U}{U_n}\right)^{-n}\left(\frac{RH_n}{RH}\right)^m \tag{10}$$

that the lifetime at 70°C and $U_n$ may be estimated to be approximatively equal to 60,000 (h) or nearly 7 years.

## 5 Technology qualification

The next thing to do is to evaluate the confidence level that, in operation, a given number of capacitors $k$ in a batch of $N$ capacitors are fulfilling the defined condition (for example, a proportion $p > 90\%$ of remaining capacitance after 100,000 h of operation).

The confidence level is given by a binomial law (failed/not failed) and will improve with the number of capacitors which pass the tests. The reverse question is: how many pieces have to be tested to get a given confidence level? The simple relation

$$n = \frac{\log(1-\text{CL})}{\log(p)} \tag{11}$$

yields the number of pieces which must tested without failure to get a confidence level CL. For example, in an accelerated test at 70°C and $1.4U_n$, $n = 16$ pieces must be tested, and none must fail during a 2000 h test, to get a confidence level of CL = 80%; in other words, 80% of all the capacitors will have a remaining capacitance greater than $p = 90\%$ in these conditions. Considering the acceleration factors, the test will state that 80% of all the capacitors will have more than 90% of their capacitance after 60,000 h of operation at $U_n$ and 40°C.

In order to introduce the new technology of oil-impregnated metallized film with an electrical field rated at 200 V/$\mu$m, the company Montena Components (today called Maxwell Technologies), in collaboration with the French railways company (SNCF), ran comparative accelerated tests during the 1990s, alongside the tests defined by the IEC 61071 standard [21]. The TGV high speed train input filter capacitor bank has a nominal voltage of 1800 V d.c. and a capacitance of 8 mF. The requested lifetime is 20 yr with a capacitance tolerance ±10%. The bank is built with four 2 mF capacitors.

The volume and weight of the capacitors vary inversely with the square of the electrical field. The first generation was operated at 150 V/μm, and each capacitor had a mass of 44 kg. In a TGV in normal commercial use, a bank at 200 V/μm (four capacitors of mass 22 kg with film thickness of 9 μm) was mounted in one of the tractors, and a bank at 240 V/μm (four capacitors of mass 17 kg with film thickness of 7.4 μm) was mounted on the other side of the train. The voltage and climatic conditions were therefore the same for the two batteries.

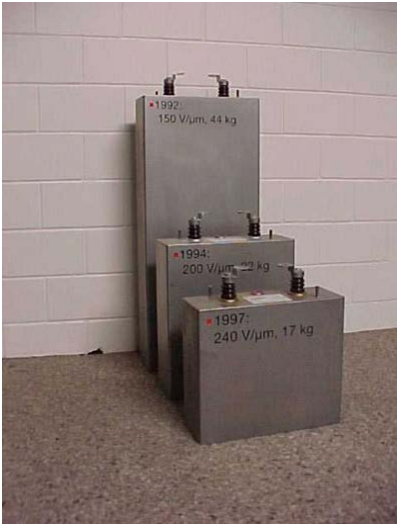

**Fig. 8:** A 2 mF 1800 V d.c. capacitor for input filters of TGV application. The sizes correspond to 150, 200, and 240 V/μm in the dielectric.

In parallel with the field experiment, a 2 mF capacitor, made with 9 μm thick film, has been tested in an oven at 70°C in the manufacturer's laboratory at 2500 V d.c., a voltage which corresponds to an electrical field of 280 V/μm. The capacitors in the TGV have been measured before the test and after 6 months and one, two, and four years, the capacitors in the laboratory have been measured weekly at the beginning and monthly at the end of the test.

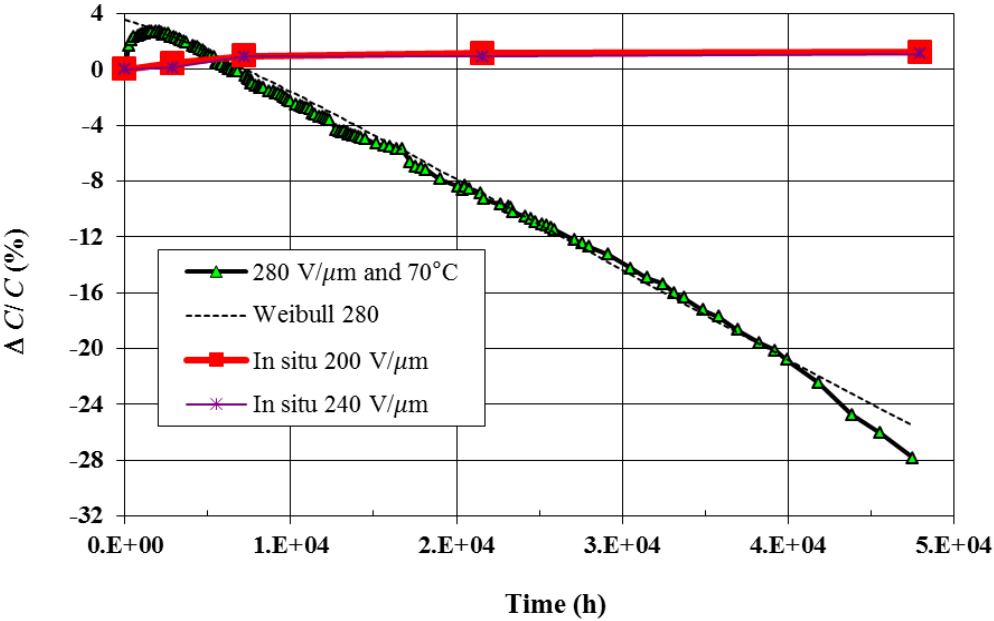

**Fig. 9:** Capacitance loss as a function of time for 280 V/μm laboratory test at 70°C and TGV in situ tests at 200 and 240 V/mm film [22].

At the beginning of the stress application a slight capacitance increase is observed because of the electrostatic compression of the films. It is also interesting to note that during four years the two capacitor batteries mounted in the TGV in commercial use are still in the compression phase.

The capacitance loss of a metallized film capacitor may be fitted by a Weibull law. In the case of an electrical field of 280 V/$\mu$m and an oven temperature of 70°C, a good fit is obtained with $p = 1.2$ and $\lambda_0 = 1/120,000$ h$^{-1}$. The 10% capacitance loss has been reached after 20,000 h at 70°C and 280 V/$\mu$m; tan $\delta$ was smaller than 70E$^{-4}$, whereas it was 30E$^{-4}$ at the beginning.

With a very rough estimation of mean temperature of 35°C inside the TGV tractor, a temperature acceleration factor of 10 between 35 and 70°C and a voltage acceleration factor of 10 between 200 and 280 V/$\mu$m, one may expect a lifetime of 2,000,000 h to reach 10% capacitance loss at 35°C and 200 V/$\mu$m. With an acceleration factor of 3 between 240 and 280 V/$\mu$m, the expected lifetime at 240 V/$\mu$m would be 'only' 600,000 h.

## 6 Conclusion

Capacitors often represent a small part of the cost of an installation. Their failure however may have huge physical and financial consequences. A typical example is a small capacitor of cost 1p connected in series in the electronic control of a freezer which leads to almost all these devices having to be scrapped when they fail. When capacitors have to be used in highly reliable applications, they should be tested in advance.